\documentclass[conference]{IEEEtran}

\usepackage[utf8x]{inputenc}
\usepackage{xcolor}
\usepackage{graphicx}
\usepackage{xspace}
\usepackage{pifont} 
\usepackage{multirow}
\usepackage{graphicx}
\usepackage{array}
\usepackage{ragged2e}
\usepackage{amsmath}
\usepackage[utf8x]{inputenc}
\usepackage{fonts-tlwg}
\usepackage{adjustbox}
\usepackage{mathptmx}
\usepackage{anyfontsize}
\usepackage{t1enc}
\usepackage{colortbl}
\let\oldding\ding
\newcommand{\newding}[2][1]{\scalebox{#1}{\oldding{#2}}}
\newcommand{\cmark}{\newding[1.3]{51}}%
\newcommand{\xmark}{\newding[1.3]{55}}%

\usepackage[thai,english]{babel}

\makeatletter
\@namedef{opt@inputenc.sty}{utf8}
\makeatother
\usepackage{CJKutf8}

\newcommand{\ignore}[1]{}

\title{On the feasibility of attacking Thai LPR systems with adversarial examples}

\begin{document}
	
	\author{\IEEEauthorblockN{Chissanupong Jiamsuchon}
		\IEEEauthorblockA{\textit{College of Computing} \\
			\textit{Prince of Songkla University}\\
			Phuket, Thailand \\
			s6230613001@phuket.psu.ac.th}
		\and
	\IEEEauthorblockN{Jakapan Suaboot}
\IEEEauthorblockA{\textit{College of Computing} \\
\textit{Prince of Songkla University}\\
Phuket, Thailand \\
jakapan.su@phuket.psu.ac.th}
		\and
		\IEEEauthorblockN{Norrathep Rattanavipanon}
		\IEEEauthorblockA{\textit{College of Computing} \\
			\textit{Prince of Songkla University}\\
			Phuket, Thailand \\
			norrathep.r@phuket.psu.ac.th}
	}
	
	\maketitle
	
	\begin{abstract}
		Recent advances in deep neural networks (DNNs) have significantly enhanced the capabilities of optical character recognition (OCR) technology, enabling its adoption to a wide range of real-world applications.
		Despite this success, DNN-based OCR is shown to be vulnerable to adversarial attacks, in which the adversary can influence the DNN model's prediction by carefully manipulating input to the model. 
		Prior work has demonstrated the security impacts of adversarial attacks on various OCR languages. However, to date, no studies have been conducted and evaluated on an OCR system tailored specifically for the Thai language. 
		To bridge this gap, this work presents a feasibility study of performing adversarial attacks on a specific Thai OCR application -- Thai License Plate Recognition (LPR). Moreover, we propose a new type of adversarial attack based on the \emph{semi-targeted} scenario and show that this scenario is highly realistic in LPR applications. Our experimental results show the feasibility of our attacks as they can be performed on a commodity computer desktop with over $90\%$ attack success rate.
		\ignore{
			Numerous real-world detection systems make use of optical character recognition (OCR). OCR could be used at some checkpoints to detect and record civilian vehicles so that the government can better manage statistics and determine the effectiveness of the system. Using an adversarial example, the adversary may find a new way to determine the threat model and fool the Thai OCR systems. Most existing adversarial attacks, however, produce adversarial examples that are overlap and heavily pollute the background. To solve this problem, we suggest a black-box semi-targeted attack that can improve Thai OCR systems' accuracy when using adversarial examples. The experimental results demonstrate the number of samples needed to determine the specific accuracy in each condition of the black-box semi-targeted attack, as well as the amount of memory needed and the amount of runtime required to execute the attack.}
	\end{abstract}
	

	\section{Introduction}

	
	Optical character recognition (OCR) is a technology to recognize characters from printed or handwritten images.
	In the last few decades, OCR has been adopted in many real-world applications mainly due to the rise of deep neural network (DNN) development.
	With DNN, OCR can now perform the character recognition task at high speed, enabling its use in many mission-critical and time-sensitive applications. For instance, an OCR system can be deployed in an airport to recognize passport information automatically~\cite{airport-ocr}; or modern license plate recognition systems employed by law enforcement rely heavily on OCR in their core engine~\cite{lpr-ocr}.
	
	Besides the timing performance, the security of OCR is also paramount to the underlying application. 
	Unfortunately, OCR inherits the same security weakness as DNN since it is also vulnerable to an attack based on \emph{adversarial examples}~\cite{goodfellow2014explaining}. 
	The aim of this attack is to confuse the DNN model, causing it to misclassify a specific input image. It is typically carried out by introducing subtle but deliberate changes to the input. These changes can be in the form of noise perturbation or small pixel images that are carefully crafted in such a way that they do not look suspicious to the human eyes. 
	As OCR has become widely adopted, it presents more incentives for an adversary to use this type of attack for his/her own benefit. This attack, for instance, can cause the OCR model to misinterpret passport data, license plate numbers, or financial documents, resulting in financial damages or crime detection avoidance.
	
	A number of prior works explore different techniques to generate adversarial examples in black-box \cite{ilyas2018black} and white-box \cite{ebrahimi2017hotflip} environments, in targeted~\cite{szegedy2013intriguing} and untargetd~\cite{moosavi2016deepfool} scenarios, and with 
	different OCR languages, e.g., English~\cite{song2018fooling}, Chinese~\cite{chen2020attacking}, and Arabic~\cite{alshemali2019adversarial}.
	Despite this rich literature, to the best of our knowledge, there has been no prior work to demonstrate the attack success on an OCR system based on \emph{Thai} language.
	Due to the idiosyncratic features of the Thai alphabet (e.g., some letters contain an upper/lower symbol -- \textthai{ฮ/ญ}), it remains unclear whether these existing attack techniques are still effective for Thai OCR systems.
	
	To this end, we set out to answer this question by demonstrating whether it is feasible to generate adversarial examples that can be used to fool the state-of-the-art Thai OCR system.
	To achieve this goal, we turn our attack focus to a specific but widely-used OCR application -- License Plate Recognition (LPR) system. 
	In particular, our attack targets an LPR system based on Google Tesseract~\cite{smith2007overview} with Thai language support.
	
	Contrary to the previous works in \cite{szegedy2013intriguing} or \cite{moosavi2016deepfool}, we consider our LPR attack scenario \emph{semi-targeted}, in which a successful adversarial example can mislead the LPR model to output any element in \emph{the set of adversary-chosen incorrect classes} (e.g., a set of valid license numbers other than the true number).
	This is distinct from the targeted scenario, which aims to misguide the model to return a \emph{particular} adversary-chosen incorrect class (e.g., a specific fake license number), or the untargeted scenario, which tricks the model into predicting \emph{any} of the incorrect classes (e.g., any sequence of Thai characters/digits other than the true license number).
	We also propose a transformation that converts the existing targeted attack into the semi-targeted attack considered in this work.
	
	Finally, we perform implementation experiments to evaluate our proposed LPR attack.
	The results indicate the realism of our attack as it obtains a high attack success rate and requires only a reasonable amount of resources (i.e., runtime and RAM usage) that can feasibly be acquired from a regular desktop computer.
	Overall, we believe this work represents the first step towards raising awareness of the threats posed by Thai OCR systems and eventually towards securing these systems against adversarial examples. 
	
	The contribution of our work can be summarized as follows:
	
	\begin{itemize}
		\item[(i)] We present a systematic approach to demonstrate the feasibility of constructing adversarial examples to fool the state-of-the-art Thai OCR-based LPR system.
		
		\item[(ii)] We explore an alternative attack scenario, called semi-targeted, and show it is highly realistic for attacking LPR applications.
		
		\item[(iii)] Our evaluation results show the feasibility of our attack; it can achieve up to 91\% attack success rate and can be carried out realistically using only a commodity computer.
	\end{itemize}
	
	\section{Background and Related Work}
	
	
	\subsection{License Plate Recognition (LPR)}
	
	LPR is the process that automatically reads and extracts vehicle license plate information from an image. It typically consists of three steps: localization, segmentation, and identification.
	In the first step, an LPR system scans through the entire image to detect and locate a license plate. 
	Then, the segmentation step extracts the regions from the detected license plate where each region contains exactly a single character. 
	Finally, LPR leverages OCR technology to classify and recognize each character and outputs the digitized license information in the identification step. 
	
	While numerous OCR techniques have been proposed for LPR systems, the most common one used by modern LPR systems is based on DNNs.
	For example, Tesseract~\cite{smith2007overview} is the state-of-the-art DNN-based OCR engine developed by Google and
	has been used in many LPR systems~\cite{palekar2017real}.
	The current version of Tesseract uses LSTM DNNs and supports more than 50 languages, including Thai.
	Besides LPR, Tesseract has been adopted to recognize Thai characters in other settings, e.g., Thai document digitization~\cite{chumwatana2021using}.

	\subsection{Adversarial Attacks}
	
	An adversarial attack was first introduced and investigated by Szegedy et al. in 2013~\cite{szegedy2013intriguing}.
	They show that by optimizing DNN's prediction error, an adversary can generate a small perturbation that can be applied to an input image in such a way that the resulting image (called \emph{an adversarial example}) is misclassified by the DNN model.
	The work in~\cite{szegedy2013intriguing} has inspired many subsequent studies to improve upon, and/or proposed different settings for, adversarial attacks. 
	Techniques in adversarial attacks can often be categorized using two orthogonal dimensions -- adversarial knowledge and goal:
	
	\begin{enumerate}
		\item \textbf{Adversarial knowledge} can be further divided into white-box and black-box environments. White-box attacks assume a powerful adversary that has complete knowledge of the DNN model's architecture, including parameters, weight values, and/or its training dataset. 
		Black-box attacks, on the other hand, consider a weaker adversary which can only query the DNN model but has no access to the model's internal information. 
		
		\item \textbf{Adversarial goal} is often classified as either targeted or untargeted scenarios. Targeted attacks aim to deceive the model into classifying an adversarial example as a targeted adversarial class, whereas an untargeted attack misleads the classification to an arbitrary class other than the correct one. 
	\end{enumerate}
	
	Prior works have explored various techniques for adversarial example generation targeting OCR systems with: 
	(i) black-box~\cite{bayram2022black} and white-box~\cite{song2018fooling} environments, 
	(ii) targeted~\cite{chen2020attacking} and untargeted~\cite{zha2020rolma} scenarios, 
	and (iii) English~\cite{song2018fooling}, Chinese~\cite{chen2020attacking}, and Arabic~\cite{alshemali2019adversarial} languages.
	In this work, we aim to assess the feasibility of performing an adversarial attack in Thai LPR systems with a realistic black-box and semi-targeted adversarial setting.

	\section{Adversary's Goal \& Threat Model}
	
	We consider a realistic adversary which aims to trick an automatic LPR system to misclassify a specific potentially illegal license plate into a different but still valid (i.e., well-formed) license number.
	The adversary is assumed to have oracle access to the black-box LPR model, i.e., he/she can query for the model's prediction output on any given image input.
	However, as the model is usually proprietary and confidential, he/she has no access to the model's internal parameters.
	
	Figure~\ref{fig:adv} shows a scenario for performing an adversarial attack on a Thai LPR system. The attack is carried out by generating an adversarial example from an illegal license plate.
	Then, it is considered a successful attack if the following requirements hold: 
	
	\begin{figure}[ht]
		\includegraphics[width=\columnwidth]{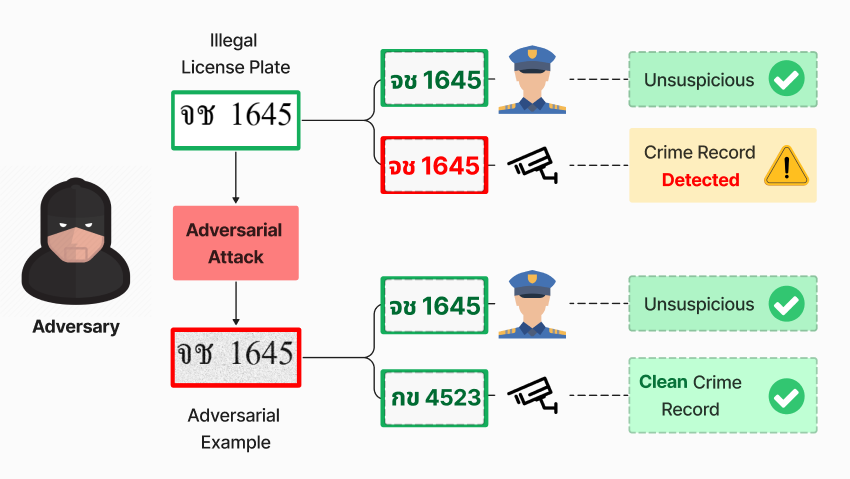}
		\centering
		\caption{Adversarial attacks on Thai LPR systems}
		\label{fig:adv}
	\end{figure}
	
	
	
	\textbf{[R1]} The generated adversarial example looks similar to the illegal license plate input in human eyes. This is to ensure that only a small change needs to be applied on the physical license plate, and as a result, 
	the modified license plate can still fool the LPR system without being noticed by humans. 
	
	\textbf{[R2]} The adversarial example's prediction class is different from its true class but still considered a \emph{valid} license number. The rationale behind this requirement is that to better evade detection, the adversary wants to avoid the DNN model returning an invalid and thus suspicious class, e.g., a malformed/unassigned license number since it can easily be detected in software or by police officers.
	
	Without loss of generality, we simplify \textbf{[R2]} by considering a license number \emph{valid} if it consists of two Thai consonants followed by a four-digit number. 
	For example, \textthai{มค3456} is valid but \textthai{มกุ1234} or \textthai{มค123} are not.
	In practice, \textbf{[R2]} can be satisfied by using a database of legal license plate numbers.
	
	Due to \textbf{[R2]}, it becomes clear that the traditional targeted and untargeted scenarios are not directly suitable in this attack setting.
	Specifically, the untargeted scenario could return an invalid number (e.g., \textthai{มค123}), violating \textbf{[R2]}; whereas the targeted scenario can be too restrictive. Hence, in this work, we introduce a relaxed concept of the targeted scenario, called \textbf{semi-targeted}, which accepts an adversarial example if its prediction class falls into a specific adversary-chosen set (as opposed to a specific class in the targeted scenario), e.g., a set of valid license numbers in the LPR application. 

	\section{Methodology}
	\label{sec:meth}
	
	
	\subsection{Overview}
	
\begin{figure}[h]
		\includegraphics[width=\columnwidth]{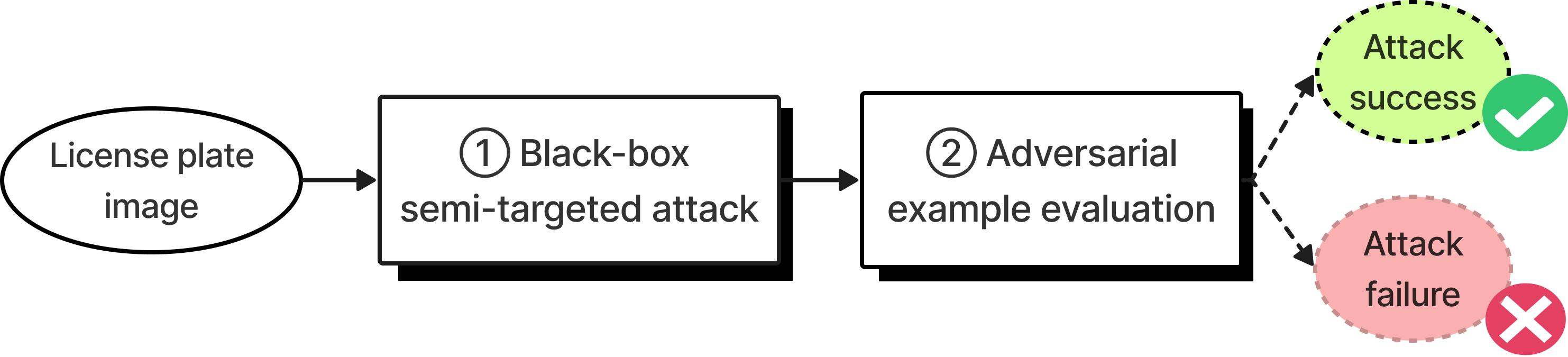}
		\centering
		\caption{Methodology for attacking Thai OCR systems}
		\label{fig:eva}
	\end{figure}

	Our methodology for attacking Thai OCR systems consists of two phases, as shown in Figure~\ref{fig:eva}. The first phase performs the black-box semi-targeted adversarial attack on an input license plate image and outputs an adversarial example. 
	
	\begin{figure*}[!htp]
		\includegraphics[width=.8\linewidth]{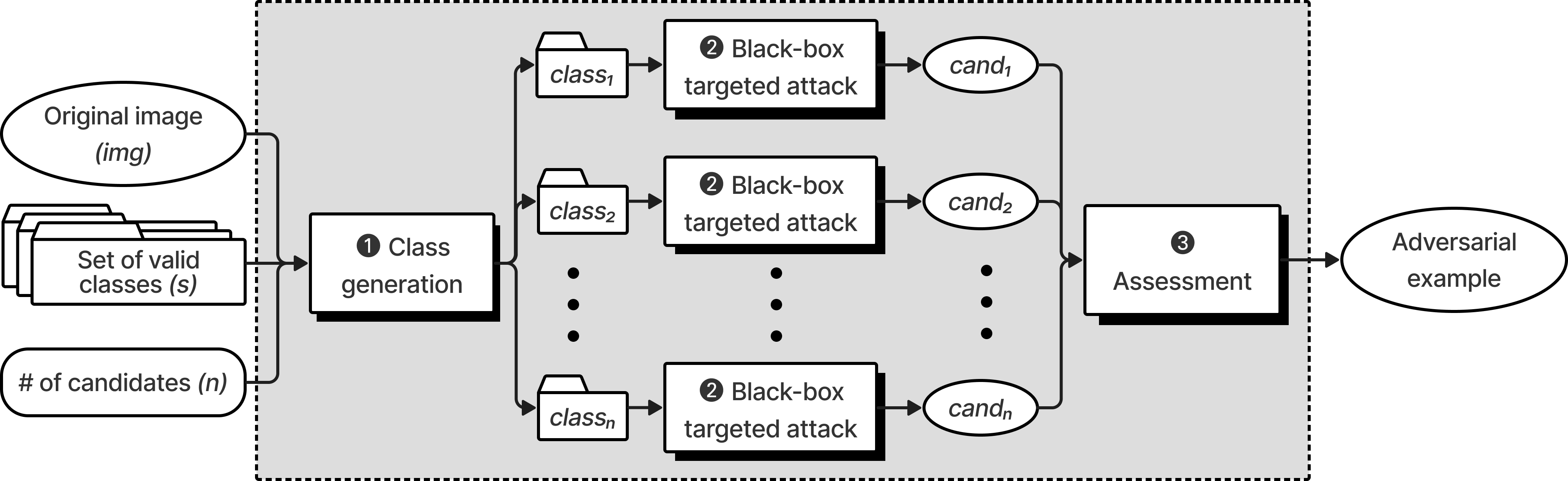}
		\centering
		\caption{Black-box semi-targeted attacks}
		\label{fig:bb}
	\end{figure*}

	The second phase takes as input, the adversarial example, and evaluates whether this adversarial example constitutes a successful attack or not.
	We now discuss each phase in detail.

	\subsection{Phase-1: Black-box Semi-targeted Adversarial Attack}
	
	As illustrated in Figure~\ref{fig:bb}, our black-box semi-targeted attack requires three input parameters: (1) an original image -- $img$; (2) a set of valid classes -- $s$; and (3) the number of candidates to be considered in this attack -- $n$. 
	In the context of LPR, $img$ represents a license plate image; $s$ corresponds to a set of valid license numbers, where, in this work, $s$ is set to common license patterns in Thailand with two Thai consonants followed by a four-digit number. 

	The attack starts in~\ding{202}. It generates $n$ classes from the given input with a constraint that all of these $n$ classes must: (1) be non-repetitive and (2) contain at least one Thai consonant different from the $img$ class. Then, we can apply the state-of-the-art black-box targeted attack for each individual class, resulting in $n$ candidates for adversarial examples in~\ding{203}.
	Finally, in~\ding{204}, we display these $n$ candidates to the user, ask the user to select the one that is closely similar to $img$, and output it as the adversarial example. 
	
	Note that this phase will always yield the adversarial example satisfying \textbf{[R2]}.
	This is because the targeted attack in~\ding{203} guarantees to produce an adversarial example that will be classified as the targeted class $class_i$, which, by construction in~\ding{202}, is valid (i.e., $class_i \in s$) and different from the $img$ class.
	
	\subsection{Phase-2: Adversarial Example Assessment}
	
	To assess the generated adversarial example, we recruit participants from our university,
	present them with the adversarial example image, and interview them with two questions:
	
	\textbf{Q1:} Are all characters legible in the presented image?
	
	\textbf{Q2:} What license number can you read from the image?
	
	The attack is considered successful if the participant responds ``yes'' to the first question and the answer from the second question matches the license number in $img$. If any of these conditions are not fulfilled, we return ``Attack failure''.
	As a result of these two carefully-crafted questions, the adversarial example can only pass this phase when still resembling $img$, thus satisfying \textbf{[R1]}.
	
	\ignore{
		Following the collection of adversarial examples, the experimenter must design a human survey in which participants are asked to evaluate the examples while being presented with two questions. 
		
		\textbf{1)} Is it possible to make out what's written inside the adversarial example?
		
		\textbf{2)} What do you understand if the writing is legible?
		Based on the answers to these two questions, we can calculate the attack success rate, \textbf{[R1]}, and the attack success rate, \textbf{[R2]} for each candidate count in the adversarial example.
	}
	
	\section{Feasibility Results}
	
	\subsection{Experimental Setup}
	
	All of our experiments were conducted on an Ubuntu 20.04 machine with an Intel i7-11700k CPU@3.60 GHz.
	To measure the attack success rate, we performed our attack on 100 unique software-generated Thai license plate images. The OCR system used in our attack was based on Tesseract v5.2.0 and ran with the following parameters: \texttt{psm=10,oem=1}. Lastly, we used HopSkipJumpAttack~\cite{chen2020hopskipjumpattack} as the underlying black-box targeted attack algorithm; for each sample, we ran this attack until it reached 300 iterations.
	
	\noindent\textbf{\underline{Ethics.}} 
	Our experiments were conducted using synthetic, instead of real, license plates for ethical reasons. This work was conducted solely for academic purposes and we do not condone using it for real-world attacks. Further, we did not gather any personally identifiable information during our interviews with participants.
	
	\subsection{Experimental Results}
	
	\begin{figure}[h]
		\includegraphics[width=.9\columnwidth]{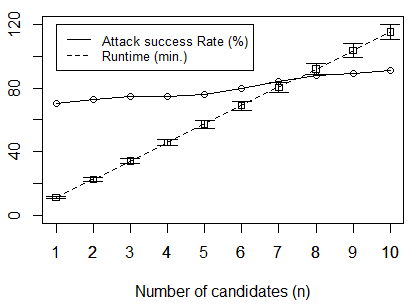}
		\centering
		\caption{Attack success rate and execution time}
		\label{fig:ATK_Runtime}
	\end{figure}
	
	\noindent\textbf{Attack Success Rate (ASR).} Figure~\ref{fig:ATK_Runtime} shows ASR of our attack while varying $n$. ASR improved drastically as we moved from the targeted attack ($n=1$) to the semi-targeted attack ($n>1$), with $ASR=91\%$ for $n=10$, compared to $ASR=70\%$ for $n=1$. This highlights the effectiveness of the semi-target scenario for attacking Thai OCR systems. We present a selection of generated adversarial examples for various $n$ values in Table~\ref{tab:result}, where Suc. refers to ``Attack success".

	\begin{table*}[!htp]
		\centering
		\caption{Samples of adversarial examples}
		\resizebox{.8\linewidth}{!}{%
			\begin{tabular}{|c|ccc|ccc|ccc|}
				\hline
				\multicolumn{1}{|c|}{\large\textbf{Sample}\cellcolor[gray]{0.8}} &
				\multicolumn{3}{c|}{\large\textbf{n=1}\cellcolor[gray]{0.8}} &
				\multicolumn{3}{c|}{\large\textbf{n=5}\cellcolor[gray]{0.8}} &
				\multicolumn{3}{c|}{\large\textbf{n=10}\cellcolor[gray]{0.8}} \\ \hline
				\textbf{Input Image}\cellcolor[gray]{0.9} &
				\multicolumn{1}{c|}{\textbf{Adv. Ex.}\cellcolor[gray]{0.9}} &
				\multicolumn{1}{c|}{\textbf{OCR Out.}\cellcolor[gray]{0.9}} &
				\multicolumn{1}{c|}{\textbf{Suc.}\cellcolor[gray]{0.9}} &
				\multicolumn{1}{c|}{\textbf{Adv. Ex.}\cellcolor[gray]{0.9}} &
				\multicolumn{1}{c|}{\textbf{OCR Out.}\cellcolor[gray]{0.9}} &
				\multicolumn{1}{c|}{\textbf{Suc.}\cellcolor[gray]{0.9}} &
				\multicolumn{1}{c|}{\textbf{Adv. Ex.}\cellcolor[gray]{0.9}} &
				\multicolumn{1}{c|}{\textbf{OCR Out.}\cellcolor[gray]{0.9}} &
				\multicolumn{1}{c|}{\textbf{Suc.}\cellcolor[gray]{0.9}} \\ \hline
				\adjustbox{valign=c}{\includegraphics[scale=0.24]{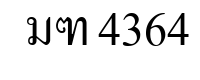}} &
				\multicolumn{1}{l|}{\adjustbox{valign=c}{\includegraphics[scale=0.24]{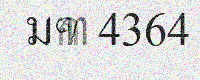}}} &
				\multicolumn{1}{c|}{\normalsize\textthai{มค}{\small 4364}} & 
				\xmark &
				\multicolumn{1}{l|}{\adjustbox{valign=c}{\includegraphics[scale=0.24]{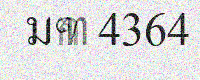}}} & 
				\multicolumn{1}{c|}{\normalsize\textthai{มค}{\small 4364}} & 
				\xmark &
				\multicolumn{1}{l|}{\adjustbox{valign=c}{\includegraphics[scale=0.24]{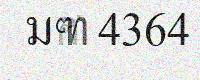}}} & 
				\multicolumn{1}{c|}{\normalsize\textthai{มศ}{\small 4364}} & 
				\xmark \\ \hline
				
				\adjustbox{valign=c}{\includegraphics[scale=0.24]{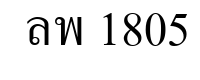}} &
				\multicolumn{1}{l|}{\adjustbox{valign=c}{\includegraphics[scale=0.24]{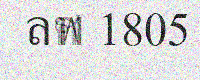}}} &
				\multicolumn{1}{c|}{\normalsize\textthai{ลศ}{\small 1805}} &
				\xmark &
				\multicolumn{1}{l|}{\adjustbox{valign=c}{\includegraphics[scale=0.24]{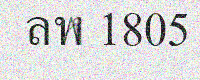}}} &
				\multicolumn{1}{c|}{\normalsize\textthai{ลห}{\small 1805}} &
				\xmark &
				\multicolumn{1}{l|}{\adjustbox{valign=c}{\includegraphics[scale=0.24]{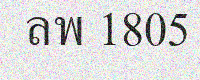}}} &
				\multicolumn{1}{c|}{\normalsize\textthai{ลม}{\small 1805}} &
				\cmark \\ \hline
				
				\ignore{
					\adjustbox{valign=c}{\includegraphics[scale=0.24]{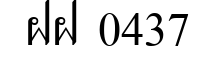}} &
					\multicolumn{1}{l|}{\adjustbox{valign=c}{\includegraphics[scale=0.24]{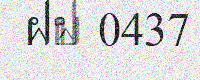}}} &
					\multicolumn{1}{c|}{\normalsize\textthai{ฝบ}{\small 0437}} &
					\xmark &
					\multicolumn{1}{l|}{\adjustbox{valign=c}{\includegraphics[scale=0.24]{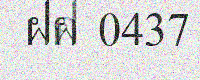}}} &
					\multicolumn{1}{c|}{\normalsize\textthai{ฝค}{\small 0437}} &
					\xmark &
					\multicolumn{1}{l|}{\adjustbox{valign=c}{\includegraphics[scale=0.24]{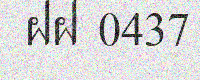}}} &
					\multicolumn{1}{c|}{\normalsize\textthai{ฝศ}{\small 0437}} &
					\cmark \\ \hline}
				
				\adjustbox{valign=c}{\includegraphics[scale=0.24]{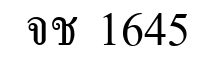}} &
				\multicolumn{1}{l|}{\adjustbox{valign=c}{\includegraphics[scale=0.24]{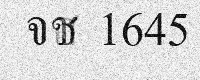}}} &
				\multicolumn{1}{c|}{\normalsize\textthai{จส}{\small 1645}} &
				\xmark &
				\multicolumn{1}{l|}{\adjustbox{valign=c}{\includegraphics[scale=0.24]{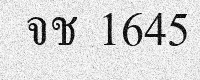}}} &
				\multicolumn{1}{c|}{\normalsize\textthai{จซ}{\small 1645}} &
				\cmark &
				\multicolumn{1}{l|}{\adjustbox{valign=c}{\includegraphics[scale=0.24]{Picture_Table/n3/Case4_n10.png}}} &
				\multicolumn{1}{c|}{\normalsize\textthai{จซ}{\small 1645}} &
				\cmark \\ \hline
				
				\adjustbox{valign=c}{\includegraphics[scale=0.24]{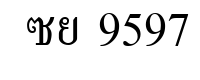}} &
				\multicolumn{1}{l|}{\adjustbox{valign=c}{\includegraphics[scale=0.24]{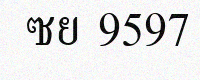}}} &
				\multicolumn{1}{c|}{\normalsize\textthai{ซฝ}{\small 9597}} &
				\cmark &
				\multicolumn{1}{l|}{\adjustbox{valign=c}{\includegraphics[scale=0.24]{Picture_Table/n1/Case2_n10.png}}} &
				\multicolumn{1}{c|}{\normalsize\textthai{ซฝ}{\small 9597}} &
				\cmark &
				\multicolumn{1}{l|}{\adjustbox{valign=c}{\includegraphics[scale=0.24]{Picture_Table/n1/Case2_n10.png}}} &
				\multicolumn{1}{c|}{\normalsize\textthai{ซฝ}{\small 9597}} &
				\cmark \\ \hline
			\end{tabular}
		}
		\label{tab:result}
	\end{table*}
	
	\noindent\textbf{Attack Resource Consumption.} In terms of resource consumption, generating adversarial examples requires a moderate amount of RAM ($\sim1.8-2$GB) on our machine, independent of the $n$ value. On the other hand, the runtime for adversarial example generation linearly depends on $n$, as shown in Figure~\ref{fig:ATK_Runtime}. 
	For $n=10$, the attack takes less than 2 hours to complete, which we consider to  be reasonable because it only needs to be done once for any given license plate.
	
	\ignore{
		Case 1 in Table 1 shows that all adversarial examples fail in every n, and it can still be seen as \textbf{[R2]} compared to Case 2, which has all passed in every n, because the adversarial examples look as good as \textbf{[R1]} at the first n means, whereas the other cases have unique results, which have textbf{[R1]} and \textbf{[R2]} mash up together.
		
		According to the overall results, the attack success rate began at 70\% and has increased with the several attack success rates in a larger number of candidates, with the last result from n = 10 having the attack success rate closed up to 91\%, which is a good accuracy that we have expected since the methodology design.
		
		Runtime execution begins with an average of 11.2814 minutes for one class per sample and an average of 115.2784 minutes for ten classes per sample. In the meantime, the runtime execution for this attack using the same environment of samples can be demonstrated in Figure 2. This means that it will increase more linearly depending on how many classes per sample we try in the experiment. It is clear that the runtime execution will increase by a factor of multiple times when there are more classes for each sample.
	}

	\section{Conclusion}
	This paper presents the first feasibility study of performing adversarial attacks on Thai OCR-based LPR systems. In addition, it proposes a new 
	type of attack scenario, called \emph{semi-targeted}, and argues that this scenario is more practical for attacking LPR systems than the traditional targeted and untargeted scenarios. Our experiments demonstrate the feasibility of our attack as it achieves a high success rate and can be carried out only using a commodity computer.
	
	
	\ignore{
		When testing the accuracy of the Thai OCR system, we found that the adversarial example attack method had the most difficulty in producing overlap characters and a noisy background, so we proposed a new methodology called a ``back-box semi-targeted attack'' to find more the feasibility by generating potential candidates. Our findings show that, depending on the pool of potential candidates, the accuracy of black-box semi-targeted attacks can range from 70\% to 91\%.}
	\bibliographystyle{plain}
	\bibliography{ref}
	
\end{document}